\documentclass[letters]{mn2e}
\usepackage{psfig}
\usepackage{mnras_cite}
\newcommand{\kel}{\mbox{ K}}

\newcommand{\da}{d_A}

\newcommand{\bn}{\hat{\bf n}}

\newcommand{\rad}{r}    

\newcommand{\Ylmn}{Y_{l}^{m}}

\begin{document}

\title[Polarization Signals of the 21 cm Background]{Polarization
Signals of the 21 cm Background from the Era of Reionization}
\author[Cooray \& Furlanetto]{Asantha Cooray and Steven
R. Furlanetto\\ Theoretical Astrophysics, Division of Physics,
Mathematics and Astronomy, California Institute of Technology \\ MS
130-33, Pasadena, CA 91125. E-mail: (asante,sfurlane)@caltech.edu}

\maketitle

\begin{abstract}
While emission and absorption lines of the 21 cm spin-flip transition
of neutral Hydrogen are intrinsically unpolarized, a magnetic field
creates left- and right-handed polarized components through the Zeeman
effect.  Here we consider the resulting polarization of the redshifted
21 cm background from the intergalactic medium before reionization.
The polarization is detectable in regions with a strong gradient in
the mean brightness temperature.  In principle, this can open a new
window on the evolution of intergalactic magnetic fields.  One
possible approach is an extended integration of an individual target
during this era, such as the Mpc-scale HII regions inferred to
surround quasars at $z \sim 6.5$.  The differential intensity between
the two polarization states can be used as a probe of the magnetic
field at the edge of the HII region.  We estimate that the SKA could
(ignoring systematics) detect $B \sim 200 \ (10) \ \mu$G coherent over
several kiloparsecs with an observational bandwidth of 100 (2)
kHz. Beyond individual sources, the statistical properties of
wide-field 21 cm polarization maps, such as the angular power
spectrum, can be used to constrain the large-scale magnetic field.  In
this case, the SKA can detect $B \sim 100 \mu$G fields coherent over
many megaparsecs.  The magnetic field can be measured in any epoch
over which the 21 cm background changes rapidly (for example because
the ionized fraction or spin temperature change).  Although the
resulting constraints with SKA are relatively weak compared to
theoretical expectations, they nevertheless offer a unique direct
probe of magnetic fields in the high-redshift universe.
\end{abstract}


\section{Introduction}
The next generation of low-frequency radio interferometers, such as
LOFAR and the SKA, is expected to open up the possibility of observing
neutral Hydrogen in the intergalactic medium (IGM) at redshifts $\sim$
10 when the universe transitions from neutral to fully ionized (e.g.,
Scott \& Rees 1990; Madau et al. 1997; Zaldarriaga et al. 2004;
Morales \& Hewitt 2004) through observations of the redshifted
spin-flip transition (with a rest wavelength of 21 cm) against the
cosmic microwave background (CMB).  This emission (during most of this
era) and absorption (at high redshifts if the IGM is cold compared to
CMB) is intrinsically unpolarized, so discussions of the 21 cm
background from the era of reionization have so far concentrated on
measuring the total intensity or brightness temperature and its
anisotropies.

While the 21 cm emission is normally unpolarized, a magnetic field can
change that.  The Zeeman effect, in general, splits the signal into
three parts: an unpolarized central component ($\pi$ component; Bolton
\& Wild 1957) bracketed in energy by two polarized components
($\sigma$ components). When viewed parallel to the magnetic field, the
central component disappears and the two $\sigma$ components are
right- and left-handed circularly polarized, respectively. The
difference in polarization states occurs because states split by the
Zeeman effect have $\pm \hbar$ momenta and, in order to conserve the
total angular momentum, the two photons must also be polarized in
opposite directions.  Polarization of the 21 cm line has been
successfully observed and used to determine the magnetic field
strength of the interstellar medium (Verschuur 1968; Verschuur 1969)
and that towards specific HI features (Verschuur 1995; Verschuur 1989)
in the Milky Way, among other applications.

Observations of the polarized 21 cm background from high redshifts
have two applications: (1) the direct determination of the magnetic
field strength in isolated regions containing neutral Hydrogen, such
as the edges of HII regions around quasars at $z \ga 6$, and (2) the
statistical determination of large-scale magnetic field properties
(such as its rms fluctuations) during the era of reionization.  We
shall see that the technique is most powerful when the mean brightness
temperature of the 21cm line changes rapidly with frequency, for
example because of sharp gradients in the ionized fraction or because
of an overall heating of the neutral gas.


This technique thus offers the potential to constrain the IGM magnetic
field on a wide range of scales.  At present, measurements of Faraday
rotation against background quasars constrains the IGM magnetic field
to be $\la 10$ nG at $z \sim 0$, if we assume a coherence length of
$\sim 1$ Mpc (see Kronberg 1994 for a review), but it could be larger
if the field were tangled on small scales.  Assuming flux
conservation, this would imply $B \la ([1+z]/7)^2 \mu{\rm G}$, if the
coherence length remained 1 Mpc comoving.  Within galaxies and galaxy
clusters, field strengths are typically $\sim 1$--$100 \mu{\rm G}$
with coherence lengths of $\la 10$ kpc (e.g., Clarke et al. 2001 and
references therein).  The origin of these fields is uncertain.
Early-universe processes can generate extremely weak seed fields on
small scales (Quashnock et al. 1989).  Fields can be amplified during
the formation of galaxies through the dynamo process (Kulsrud 1999),
but such a process will not work on the large scales of galaxy
clusters (or the IGM).  Seed fields could also be generated during
reionization itself (Gnedin, Ferrara, \& Zweibel 2000).  A more
promising possibility is for quasar or supernova winds to entrain the
fields and carry them into the IGM (Kronberg et al. 1999, 2001;
Furlanetto \& Loeb 2001; Gopal-Krishna \& Wiita 2001).  Measurements
of the field strength at $z \ga 6$ could offer invaluable insight into
these processes: because structure formation is much less advanced,
galaxies and quasars should have polluted much less of the universe.
However, placing such constraints promises to be difficult.  For
example, the number of bright sources available for Faraday rotation
observations is probably quite small, and the rotation may be
dominated by the low-redshift universe because of the much larger path
length.  The polarized 21 cm background presents an alternate
possibility; although we shall see that the constraints are fairly
weak compared to most magnetic field generation scenarios, they are
still the most promising direct measurements for the high-redshift
universe.

The discussion is organized as follows: in \S~2, we discuss the
polarized 21 cm signal and consider two separate applications
involving individual targets (\S~2.1) and large-scale statistics
(\S~2.2).  In the same sections, we also discuss the potential
detectability of these signatures with upcoming low-frequency radio
telescopes arrays such as the SKA. We conclude with a summary in \S~3.
Throughout the paper, we make use of the WMAP-favored LCDM
cosmological model (Spergel et al. 2003).

\section{Method of Calculation}

In the presence of a magnetic field, the Zeeman effect causes a level
splitting of the 21 cm line with magnitude $\delta \nu \equiv |\nu -
\nu_0|$ (Lang 1999) such that
\begin{eqnarray}
\nu &=& \nu_0 \pm  \frac{eB}{4\pi m_e c} \nonumber \\
 &=& 1420\; {\rm MHz} \pm 14 {\rm Hz} \; \left(\frac{B}{10\; \mu{\rm
 G}}\right) \, . 
\end{eqnarray}
For typical magnetic fields of order $\sim 10 \mu{\rm G}$, the
fractional separation of the two states is significantly smaller than
the line width of the 21 cm emission (typically many kHz or
more).  Thus, observing two separate peaks in the line profile is not
possible. The polarization of these two states, however, is a
potentially measurable quantity if the magnetic field strength is such
that the magnitude of the difference between left- and right-handed
polarization is larger than the detector noise.

We denote the magnetic field parallel to the line of sight as
$B_\parallel = \bn \cdot {\bf B}$.  We then write the total brightness
temperature and the difference in the brightness temperature between
left- and right-handed polarization maps as:
\begin{eqnarray}
\Delta T_{\rm tot}(\nu) &=& \Delta T_L(\nu) + \Delta T_R(\nu) \nonumber \\
\Delta P_{\rm diff}(\nu) &=& |\Delta T_L(\nu)-\Delta T_R(\nu)| \, .
\end{eqnarray}
The difference term can be simplified because $\Delta
T_L(\nu) \propto \Delta T_{\rm tot}(\nu_0-\delta \nu)$ while $\Delta
T_R(\nu) \propto \Delta T_{\rm tot}(\nu_0+\delta \nu)$, as the two
states separated in frequency are also polarized in opposite
directions.  Thus
\begin{eqnarray}
\Delta P_{\rm diff}(\nu) &=& 2 \frac{d\Delta T_{\rm tot}}{d\nu}(\nu)
\; \delta \nu \nonumber \\ 
&=& 28 \Delta T^{'}_{\rm tot}(\nu) \left(\frac{\bn \cdot {\bf B}}{10\;
  \mu{\rm G}}\right) \, , 
\end{eqnarray}
where $\Delta T^{'}_{\rm tot}(\nu)$ is the derivative of the emission
profile with respect to the frequency (in Hz).  Polarization
observations have the advantage that, although $\delta \nu$ is
extremely small, one can still constrain magnetic field with an
observed bandwidth significantly larger than $\delta \nu$, so long as
the gradient $\Delta T^{'}_{\rm tot}$ over that frequency range is
large.

During the era of reionization, there are two distinct possibilities
for polarization observations: (a) individual targets, such as HII
regions surrounding $z \sim$ 6 quasars, and (b) maps of the high-$z$
universe as a whole.  We will first discuss the extent to which
magnetic fields in individual quasar regions can be measured with
first-generation 21 cm interferometers, such as LOFAR and the SKA and
then discuss applications of statistical studies with wide-field maps
of the 21 cm background.

\subsection{Individual HII Regions}

It is now widely believed that the reionization process was both
inhomogeneous and patchy on large scales (Barkana \& Loeb 2001;
Furlanetto, Zaldarriaga, \& Hernquist 2004a). Luminous sources such as
the first galaxies and quasars ionized their surroundings and these
ionized patches grew and overlapped until they completely ionized the
universe. These ionized patches contribute to CMB temperature (Santos
et al. 2003) and polarization anisotropies (Zaldarriaga 1997), cause
fluctuations in the 21 cm background (Furlanetto, Zaldarriaga, \&
Hernquist 2004b), and correlate the two (Cooray 2004a,b). The
reionization process was not instantaneous (Wyithe \& Loeb 2003; Cen
2003; Chen et al. 2003; Haiman \& Holder 2003) and could, in certain
models, explain both the high electron scattering optical depth
$\tau_{\rm es} = 0.17 \pm 0.04$ measured through the large angular
scale bump in the CMB temperature-polarization cross-correlation
(Kogut et al. 2003) and the rapidly evolving neutral fraction $x_H$ at
$z\sim$ 6 towards Sloan Digital Sky Survey quasars (Fan et
al. 2002). During the partially ionized phase, each high-$z$ quasar should
be surrounded by an HII region expanding into a largely neutral
medium, with its size determined by the quasar luminosity and spectrum
(Wyithe \& Loeb 2004b, Mesinger \& Haiman 2004).

While 21 cm intensity observations of these sources could reveal the
shape of these HII regions, especially in combination with Ly$\alpha$
optical depth profiles (Wyithe \& Loeb 2004b), polarization
measurements can reveal the presence of a magnetic field.  To estimate
the expected polarization signal, we only need the width of the edge
of the HII region (which determines $\Delta T'_{\rm tot}(\nu)$ in
eq. [3]).  This is determined by the mean free path $\lambda$ of an
ionizing photon, which in turn depends on the photon energy (and hence
the quasar spectrum).  Using a typical quasar spectrum, Wyithe \& Loeb
(2004b) estimate $\lambda \sim 1.5 x_H^{-1} (1+z)^{-3}$ Mpc (proper).
This corresponds to a frequency width of $\Delta \nu_{\rm HII} \sim 2$
kHz, much smaller than the resolution of any of the upcoming
experiments.  This is only a rough estimate given uncertainties in the
exact shape of the quasar UV spectrum and the physical state of the
IGM surrounding these quasars, but it suffices for simple estimates of
the signal.

Using the above case as an example, we estimate polarized signals of
order $\Delta P_{\rm diff} \sim 0.3 {\rm mK} (B_\parallel/10 \mu {\rm
G})$ if the observed bandwidth $\Delta \nu_{\rm ch}$ exceeds $\Delta
\nu_{\rm HII}$.  Of course, only the transition region produces
polarized flux, so the fractional polarization is $\sim (\Delta
\nu_{\rm HII}/\Delta \nu_{\rm ch}) (\Delta P_{\rm diff}/\Delta T_{\rm
tot}) \sim 0.03\%$ for $\nu_{\rm ch}=0.1$ MHz.  
Note that this assumes
a field coherent across the several kiloparsec thickness of the
transition region; field reversals (either in frequency space or
across the telescope beam) would cause the net polarization to execute
a random walk and weaken the limit.  This weak signal presents a
substantial challenge for an unambiguous detection given the large
number of systematics in these measurements.  Nevertheless, with the
rapid improvement in technology and the unknown astrophysical
environment (which could potentially increase the amplitude of
polarized components), it is useful to estimate the level of magnetic
fields detectable by future instruments.

We let $N_{\Delta \nu_{\rm ch}}$ be the rms noise in each polarization
component over a bandwidth $\Delta \nu_{\rm ch}$.  The minimum
detectable magnetic field strength (parallel to the line of sight), at
the signal-to-noise level of unity, is then
\begin{equation}
B^{\rm min}_\parallel \; (\mu{\rm G})= \frac{\sqrt{2} N_{\Delta
    \nu_{\rm ch}}}{2.8 \Delta T^{'}_{\rm tot}(\nu)[\Delta
\nu_{\rm HII}/\Delta \nu_{\rm ch}]}\, . 
\end{equation}
The factor $\sqrt{2}$ accounts for the fact that the measurement is a
difference between polarizations.  
The bandwidth factor appears
because the signal contributes to only a fraction of the channel (we
have assumed $\Delta \nu_{\rm ch} \ge \Delta \nu_{\rm HII}$ here).
The root mean square noise contribution in a single
SKA\footnote{http://www.skatelescope.org for details} channel is
\begin{eqnarray}
N_{\Delta \nu} &=& 
0.079 {\rm mK} 
\left(\frac{203 {\rm MHz}}{\nu}\right)^2 
\left(\frac{A_{\rm eff}/T_{\rm sys}}{2 \times 10^4 {\rm m^2
    K^-1}}\right)^{-1} \nonumber \\  
&\times& \left(\frac{4'}{\Delta \theta}\right)^2 
\left(\frac{0.1 {\rm MHz}}{\Delta \nu_{\rm ch}} \ \frac{30 {\rm
    days}}{t_{\rm int}}\right)^{1/2} \, ,\nonumber \\
\end{eqnarray}
where $A_{\rm eff}/T_{\rm sys}$ captures the sensitivity of an array
with a collecting area of $A_{\rm eff}$ and a system temperature
$T_{\rm sys}$, $\Delta \theta$ is the beam size, and $t_{\rm int}$ is
the total integration time. The parameters are for the straw-man SKA,
with revised values scaled from the estimate in Zaldarriaga et
al. (2004).  In one month of continuous integration, SKA can reach
noise levels of $\sim 80$ $\mu$K with these parameters, so one could
potentially limit the magnetic field to a level of $\sim$ 200 $\mu$G
at the one-sigma level.  If we reduce the bandwidth to $\Delta
\nu_{\rm ch}=10 \ (2)$ kHz centered on the edge of the HII region, the
one-sigma constraint is $\sim$ 60 (9) $\mu$G. While smaller bandwidths
are desirable, resulting increase in systematics complicate such an 
observation.  Narrow channels of order 2 kHz are also limited
by the curvature of the HII region across the beam (which would cause
the edge to move in frequency space).  The latter could be avoided by
careful binning of smaller pixels.

Thus, for a reasonable bandwidth, the expected SKA limit is
significantly higher than many of the theoretical expectations at
these redshifts.  Models in which quasar outflows or starburst winds
pollute the IGM usually have an average field strength of $\sim 1
\mu$G in the magnetized regions (e.g.  Furlanetto \& Loeb 2001) and
primordial magnetic fields are expected to be much weaker.  Reaching a
$\sim$ 1 $\mu$G detection level would require an order of magnitude
increase increase in sensitivity relative to SKA and a better control
of systematics such that observations can be made over a bandwidth of
a few kHz level.  However, it is important to note that strengths of
$\sim 1 \mu$G are only average levels.  The fields could vary
spatially within magnetized regions, especially near galaxies and
quasars.  Because the edge of the HII regions are several kpc
thick, this test would be sensitive to any strong fields tangled on
these scales.

\subsection{Anisotropy Studies}

In addition to individual targets, one can also measure polarization
statistics with wide-field maps. This is analogous to proposed
measurements of the 21 cm power spectrum (Zaldarriaga et al. 2004),
and we take a similar approach to describing the polarized
visibilities.

In the case of polarization, we write the multipole moments of a
difference map of the two orthogonal states as
\begin{equation}
a_{lm}^{\rm diff} = \int d\bn \Delta P_{\rm diff}(\bn) \Ylmn {}^*(\bn) \, ,
\end{equation}
where $\Delta P_{\rm diff}(\bn)$ is now the difference in maps when
observations have integrated over a finite bandwidth such that
\begin{equation}
\Delta P_{\rm diff}(\bn) = 2.8 \int d\rad \Delta T^{'}_{\rm tot}(\bn \rad, \rad) \left[\bn \cdot {\bf B}(\bn \rad,\rad)\right]
W_\nu(\rad) \,.
\end{equation}
Here the magnetic field is measured in units of $\mu$G, $r$ is the
comoving distance, and $W_\nu(\rad)$ describes the experimental
response, assumed to peak at a distance $r_0$.  The window function
thus picks out a particular redshift interval because the 21 cm
feature is a spectral line.  We take the response function to be a
Gaussian of width $\Delta \nu_{\rm ch}$.  The frequency derivative can
also be transformed to a function of redshift via $d \Delta T(z)/d\nu
= -\nu_0/\nu^2 d \Delta T(z)/dz$, where $\nu$ is the observed
frequency, $\nu = \nu_0(1+z)^{-1}$.

The maximum magnitude is set by the overall strength of the 21 cm
intensity.  Following Zaldarriaga et al. (2004), the mean brightness
temperature of the 21 cm line is
\begin{eqnarray}
&&\Delta T^{\rm 21cm}(z) = \nonumber \\
&& {23.2 \; \rm mk} \left( \frac{\Omega_b h^2}{0.02}
  \right)x_H(z) \left[\frac{T_S-T_{\rm CMB}}{T_S} \right]
  \sqrt{1+z} \, , 
\label{eq:dt}
\end{eqnarray}
where $T_{\rm CMB} = 2.73 (1+z) \kel$ is the CMB temperature at
redshift $z$ and $T_S$ is the gas spin temperature.  Henceforth for
simplicity we will neglect spatial fluctuations in $T_S$ and $x_H$
(though we will allow both to be functions of redshift).  The angular
power spectrum of the polarization difference map includes
fluctuations both in the neutral Hydrogen density field (which are
captured by eq. [\ref{eq:dt}]) and spatial fluctuations of the
magnetic field.  With our approximation, fluctuations in the density
cause $\delta T(\bn) = \Delta T^{\rm 21cm}(\bn)[1+\delta_g(\bn)]$,
where $n_{H} = \bar{x}_H \bar{n}_g(1+\delta_g)$, the mean baryonic
density of the universe is $\bar{n}_g$, and the mean neutral fraction
is $\bar{x}_H$.  They can thus be encapsulated in $P_{gg}(k)$, the
spatial power spectrum of gas density fluctuations (with respect to
the mean density).  The remaining term depends on $P_{BB}(k)$, the
spatial power spectrum of the magnetic field (which we will measure in
absolute units of $\mu$G$^2$ Mpc$^{-3}$).  The relation between the
two power spectra is currently unknown.  It is possible that the
coherence scale of the magnetic field is larger than that of the
density field (this is similar to the the velocity field, which has a
larger correlation length than the density field).  On the other hand,
the opposite situation could apply if magnetic fields are highly
tangled.

In either case, the power spectrum may then be written
\begin{equation}
C_l^{\rm diff} =  \int \frac{d\rad}{d_A^2} \left[2.8 \frac{d\Delta
    T^{\rm 21cm}}{dz} \frac{\nu_0}{\nu^2}\right]^2  
I\left(\frac{l}{\da};\rad\right)  W_\nu(\rad)^2 \, 
\end{equation}
in the Limber approximation appropriate for a flat sky (Limber 1954).
Here the mode coupling integral $I$ is a convolution between
fluctuations in the gas density field and the magnetic field, weighted
by a geometric term involving the line of sight angle,
\begin{equation}
I(k) = \int k_1^2 \frac{dk_1}{8\pi^2} \int_{-1}^{+1} d\mu
\frac{(1-\mu^2)}{y_2^2} P_{gg}(ky_2)P_{BB}(ky_1) \, ,
\end{equation}
where $y_1=k_1/k$ and $y_2=k_2/k=\sqrt{1-2\mu y_1 +y_1^2}$.  This form
for the power spectrum occurs because equation~(7) presents a
convolution between the frequency derivative of the mean brightness
temperature and fluctuations in the line-of-sight magnetic field. The
same formula can also be derived as an overall modulation of the mean
brightness temperature by fluctuations in the magnetic field and, in
this sense, the derivation is analogous to calculations in the
literature (e.g., Cooray 2001) of the Ostriker \& Vishniac (1987)
effect, which is a modulation of the density field by the velocity
fluctuations.

In the limit in which small-scale fluctuations in the brightness
temperature (from density structure) modulate a more coherent magnetic
field, $k_2 = |{\bf k}-{\bf k_1}| \sim k$ such that $y_1 \ll 1$ and
$y_2 \rightarrow 1$.  Thus, $I(k) \rightarrow P_{gg}(k) \int k_1^2
dk_1/2\pi^2 P_{BB}(k_1) \int_{-1}^{+1} d\mu (1-\mu^2)/4$.  The
integral over $\mu$ simplifies to $1/3$ such that $I(k) \rightarrow
P_{gg}(k) B^2_{\rm rms}(\rad)/3$, where 
\begin{equation}
B^2_{\rm rms} = \int \frac{k^2 dk}{2\pi^2} P_{BB}(k) \, .
\end{equation}
Thus, if the magnetic field is spatially uniform on 
larger scales than the density fluctuations, we have
\begin{equation}
C_l^{\rm diff} =  \int \frac{d\rad}{d_A^2} \left[2.8 \frac{d\Delta
    T^{\rm 21cm}}{d\nu}W_\nu(\rad)\right]^2  
P_{gg}\left(\frac{l}{\da};\rad\right) \frac{B^2_{\rm rms}(\rad)}{3}.
\label{eq:cl}
\end{equation}
In this case, the factor $1/3 B^2)_{\rm rms}$ can also be understood
as a random averaging when the two line of sight angles are nearly
parallel (i.e. the small angle approximation):
\begin{equation}
\langle [\bn \cdot {\bf B}(\rad_1)][\bn' \cdot {\bf B}(\rad_2)]
\rangle \approx \frac{1}{3} \langle B^2 \rangle \, . 
\end{equation}

\begin{figure}
\centerline{\psfig{file=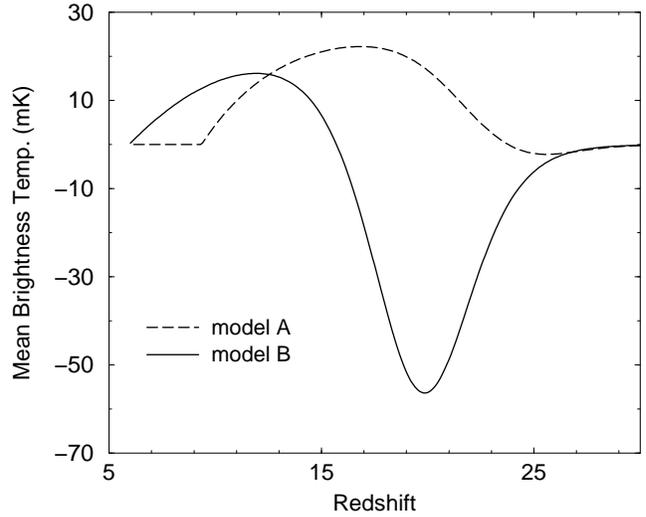,width=3.3in,angle=-90}}
\caption{Mean brightness temperature as a function of the redshift for
two different models of the reionization history.  In model A (dashed
line), we have $\zeta=20$, $f_\star=0.05$, and $f_{\rm heat}=1$.  Here
heating occurs early and $\Delta T^{\rm 21 cm}$ traces $\bar{x}_H$ for
$z \la 15$.  Model B (solid line) has $\zeta=6$, $f_\star=0.03$, and
$f_{\rm heat}=0.01$.  Heating is much less rapid in this case, but
$\Delta T^{\rm 21 cm}$ still traces $\bar{x}_H$ for $z \la 11$.}
\label{fig:dt}
\end{figure} 

We will now restrict ourselves to the simple limiting case of
eq. (\ref{eq:cl}) appropriate for large-scale magnetic fields.  To
proceed, we need a model for $d\Delta T/dz$.  This has three
components: the cosmological density evolution, $\bar{x}_H$, and
$T_S$.  We compute $\bar{x}_H$ following Furlanetto et al. (2004a),
who assume that the ionized fraction is proportional to the collapse
fraction (with a constant $\zeta$, part of which is determined by the
star formation efficiency $f_\star$).  The spin temperature evolution
is more subtle.  At these redshifts, $T_S$ is coupled to the gas
kinetic temperature by Ly$\alpha$ photons (Field 1959; Madau et
al. 1997).  We follow this background in the same way as Ciardi \&
Madau (2003), assuming that five photons are produced near Ly$\alpha$
for every ionizing photon, as appropriate for Population II stars with
a constant star formation (see their Fig. 1).  We assume that the IGM
cools adiabatically after decoupling from the CMB.  We then heat it
through X-rays from supernovae inside the ionizing sources.  We assume
one supernova occurs per 100 $M_\odot$ of stars and that each has
$10^{51}$ ergs.  We further assume that a fraction $f_{\rm heat}$ of
the supernova energy escapes into X-rays and uniformly heats the IGM.
Fig.~1 illustrates two models.  Model A is a `maximal heating'
scenario in which $\Delta T^{\rm 21cm}$ rises rapidly and decreases
later only through reionization.  Model B is a `minimal heating'
scenario in which the IGM remains cool for a long period but still
emits well before reionization occurs.  Neither of these models can be
taken as more than a representative possibility, as they neglect a
number of important effects (such as shock heating in the IGM, other
X-ray sources, and a rigorous treatment of the Ly$\alpha$ background),
but they suffice for the purposes of estimating the brightness
temperature gradients for our calculation.

In Fig.~2, we show the polarization and temperature angular power
spectra for model B at $z=9$. In these calculations, we describe the
density power spectrum with the linear density field, but scaled by a
bias factor appropriate to the distribution of neutral Hydrogen (Mo et
al. 1997).  Specifically, we follow the methods of Santos et
al. (2003, 2004), who use an approach similar to halo models of large
scale structure (Cooray \& Sheth 2002). We also show noise curves for
the SKA and for an `extended SKA'.  The polarized signal is
proportional to the brightness temperature gradient.  We have chosen a
point at which this is relatively shallow in Fig.~1; if the rapid
temperature evolution of model B at $z \sim 15$--$25$ is accessible to
observations, the polarized signal could be a few times larger.  
The signal for the late stages of reionization in model A would be
comparable to that shown in Fig.~2.  Moreover, we have assumed a
uniform $\bar{x}_H$; spatial variations in the ionized fraction (or in
$T_S$) could also increase the signal by a factor of a few.  On the
other hand, we have neglected possible correlations between the
density and the magnetic field, and we have neglected small-scale
structure in $B$.  Our estimate can therefore only be taken as a guide
to the expected order of magnitude of the signal.

We can estimate the smallest detectable $B_{\rm rms}$ using the Fisher
matrix approach. Since only a single parameter is involved, we can
write the limit on the magnetic field as simply
\begin{equation}
\sigma^{-2}(B_{\rm rms}) = \sum_l \frac{1}{(\sigma_l^{\rm
    diff})^2}\left(\frac{\partial C_l^{\rm diff}}{\partial B_{\rm
    rms}}\right)^2 \, , 
\end{equation}
where 
\begin{equation}
 \sigma_l = \sqrt{\frac{2}{(2l+1)f_{\rm sky}}} \left(C_l^{\rm diff} +
 C_l^{\rm noise}\right) \, , 
\end{equation}
with $f_{\rm sky}$ the fraction of sky observed and $C_l^{\rm noise}$
the instrumental noise.  We make the null hypothesis $C_l^{\rm
diff}=0$ and estimate the one-sigma constraint on $B_{\rm rms}$ for
the two different noise curves shown in Fig.~2. For the standard SKA
noise parameters with a one month observation), we find $B_{\rm rms}
\geq 224 (f_{\rm sky}/0.1)^{-0.5}$ $\mu$G is detectable (at
$1\sigma$). With an order of magnitude improved sensitivity and a
year-long observation, the limit improves to $B_{\rm rms} \geq 7.5
(f_{\rm sky}/0.1)^{-0.5}$ $\mu$G.  As with targeted observations of
HII regions, it is thus difficult to probe the $\mu$G fields expected
from many models.  Long integrations with instruments beyond the SKA
will be required to reach the necessary limits.

\begin{figure}
\centerline{\psfig{file=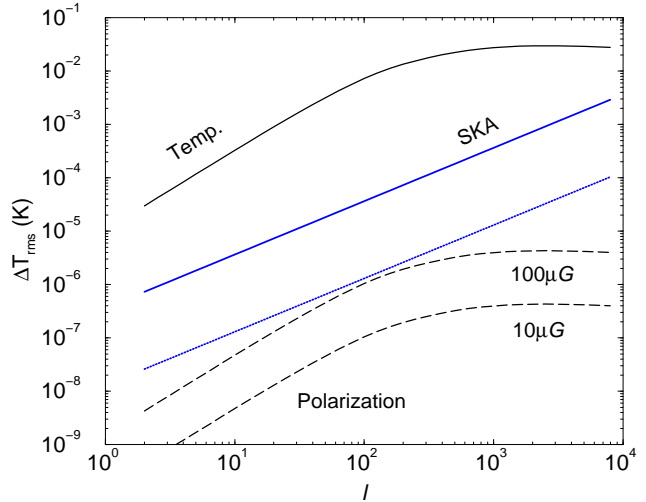,width=3.3in,angle=-90}}
\caption{The rms brightness temperature fluctuations of the 21 cm
background at $z=9$ ($\nu_{\rm obs}$ = 142 MHz), where $\Delta T_{\rm
rms} = \sqrt{l(l+1)C_l/2\pi}$. The top solid curve (labeled 'Temp.')
shows the total temperature fluctuations, while the long-dashed curves
show the fluctuations in the polarization difference assuming a large
scale magnetic field with $B_{\rm rms} = 10$ and $100 \mu$G.  Both
assume model B of Fig.~1.  The `SKA' curve shows the instrumental
noise, unbinned in the multipole space, for a one month integration
with $\Delta \nu_{\rm ch}=0.5 MHz$. The bottom (linear) curve assumes
an order of magnitude improvement in sensitivity over the SKA, a
year-long integration, and a total bandwidth of 8 MHz.}
\label{fig:cl}
\end{figure}

\section{Summary}

While emission (and absorption) from the 21 cm spin-flip transition of
neutral Hydrogen is intrinsically unpolarized, a magnetic field
creates left- and right-handed polarization through the Zeeman
effect. For individual targets of neutral Hydrogen during the era of
reionization, such as Mpc-scale HII regions surrounding quasars at $z
\ga 6$, the difference in intensity between these two polarization
states can be used as a probe of the magnetic field on the scale of
the edge of the HII region.  For example, observations of the known
SDSS quasars at $z \sim 6.5$ with the SKA can constrain magnetic
fields to a level of 200 $\mu$G (at the $1\sigma$ level) in an
integration of one month over a bandwidth of 100 kHz. If a few kHz
bandwidth observations can be made with systematics well under
control, the limit improves to $\sim$ 10 $\mu$G, again at the
$1\sigma$ level.  The expected fields have $B \la 3 \mu$G so one could
potentially reach an interesting constraint through a long integration
or an even larger instrument.  In addition, these measurements are
sensitive to fields with scales of several kpc, which could be much
larger than the mean value.

Beyond individual sources, the statistical properties of 21 cm
polarization in wide-field maps, such as the angular power spectrum of
fluctuations, can also be used to constrain large scale fluctuations
of the magnetic field. This requires the mean brightness temperature
of the 21 cm background to vary rapidly as a function of the
frequency.  Thus polarization detections are limited to epochs over
which the underlying properties of the 21 cm background, such as the
ionized fraction and spin temperature, change rapidly.  Using simple
models for the evolution of the mean brightness temperature, we have
estimated the statistical constraints that can be placed on a magnetic
field coherent over scales larger than that of the density field for
such a measurement.  The SKA can detect a large-scale field with
$B_{\rm rms} \geq 224 (f_{\rm sky}/0.1)^{-0.5}$ $\mu$G. A future
instrument with an order of magnitude better sensitivity can improve
the limit to $B_{\rm rms} \geq 7.5 (f_{\rm sky}/0.1)^{-0.5}$ $\mu$G
with a year-long observation.  Constraints of this order would be
desirable given that galaxy clusters at $z \la 1$ are known to have
$B_{\rm rms} \sim 10$--$100 \mu$G (e.g., Clarke et al. 2001).

Finally, we emphasize that, although the limits obtainable from the 21
cm background are weak compared to theoretical expectations, obtaining
other direct constraints on the magnetic field during the era of
reionization promises to be equally challenging.  This technique may
offer the first direct limits on high-redshift magnetic fields.

{\it Acknowledgments:} This work was supported at Caltech by a Senior
Research Fellowship from the Sherman Fairchild foundation.  We thank
John Kovac for useful discussions.

\end{document}